# Prediction of Lung CT Scores of Systemic Sclerosis by Cascaded Regression Neural Networks

Jingnan Jia [a,b], Marius Staring [a,b], Irene Hernández-Girón [a,b], Lucia J.M. Kroft [b], Anne A. Schouffoer [c], Berend C. Stoel* [a,b]

[a]Division of Image Processing, [b]Department of Radiology, [c]Department of Rheumatology, Leiden University Medical Center (LUMC), P.O. Box 9600, 2300 RC, Leiden, The Netherlands.

## ABSTRACT

Visually scoring lung involvement in systemic sclerosis from CT scans plays an important role in monitoring progression, but its labor intensiveness hinders practical application. We proposed, therefore, an automatic scoring framework that consists of two cascaded deep regression neural networks. The first (3D) network aims to predict the craniocaudal position of five anatomically defined scoring levels on the 3D CT scans. The second (2D) network receives the resulting 2D axial slices and predicts the scores. We used 227 3D CT scans to train and validate the first network, and the resulting 1135 axial slices were used in the second network. Two experts scored independently a subset of data to obtain intra- and inter-observer variabilities and the ground truth for all data was obtained in consensus. To alleviate the unbalance in training labels in the second network, we introduced a sampling technique and to increase the diversity of the training samples synthetic data was generated, mimicking ground glass and reticulation patterns. The 4-fold cross validation showed that our proposed network achieved an average MAE of 5.90, 4.66 and 4.49, weighted kappa of 0.66, 0.58 and 0.65 for total score (TOT), ground glass (GG) and reticular pattern (RET), respectively. Our network performed slightly worse than the best experts on TOT and GG prediction but it has competitive performance on RET prediction and has the potential to be an objective alternative for the visual scoring of SSc in CT thorax studies.

## 1. INTRODUCTION

Systemic sclerosis (SSc) is a rare immune-mediated connective tissue disease that affects different organs, with high mortality. Pulmonary disease is the leading cause of mortality in patients with SSc[1]. Because CT scans can provide accurate information on the lung, Goh[1] proposed a sensitive scoring system to quantify the extent of SSc disease from CT scans, further standardized by a CT reference atlas[2]. In this so-called Goh scoring system, 2D CT images are scored at five levels: 1) origin of the great vessels; 2) main carina; 3) pulmonary venous confluence; 4) halfway between the third and fifth level; 5) immediately above the right hemi-diaphragm. At each level, three patterns are scored as the percentage of the total lung area: total disease extent (TOT), ground-glass opacities (GG), and reticular patterns (RET). TOT is a powerful predictor[1] for mortality of SSc[1], GG is a significant biomarker to differentiate SSc and idiopathic pulmonary fibrosis[3], and RET is a strong determinant of a decline in the forced vital capacity and progression-free survival of SSc[1]. Therefore, these three patterns are important markers to monitor disease progression.

The Goh scoring system is, however, laborious, subjective and dependent on rater experience. A great number of similar scoring systems also met such challenges[4]. Therefore, an automatic scoring tool would overcome these limitations and benefit clinical application. Although attempts have been made to automatically predict biomarkers[5], their common drawback is that they use segmentation networks to output the pixel labels as a basis for computing the final biomarkers, which requires time-consuming and laborious manual pixel-wise annotations as reference data.

The purpose of this paper is to build an automatic framework to identify the five levels and subsequently score the extent of disease from CT scans, without the need for segmentation annotations. A data synthesis technique for the disease patterns (ground-glass opacities and reticulation) was introduced to alleviate the lack of training data and label unbalance, which improved the performance significantly.

## 2. METHODOLOGY

The proposed two-step framework is shown in Figure 1. In the first step, a *level selection network* (3D VGG11[6]) identifies five levels from the input 3D CT scans. Subsequently, a *Goh score prediction network* (2D VGG11[6]) predicts three scores (TOT, GG, and RET) for each input 2D slice. The pre-trained weights from ImageNet were applied to the 2D VGG11. The loss function for both networks was the mean squared error (MSE) with the manually annotated levels and the manual Goh scores, respectively. Datasets and the methodology details of two networks are described below.

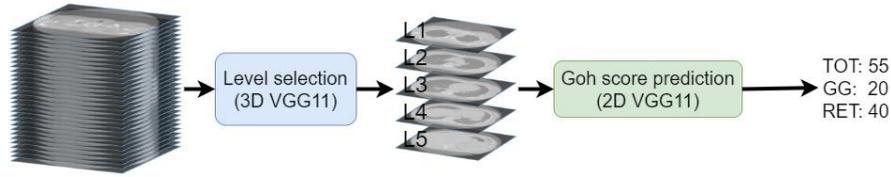

Figure 1. Proposed framework. 3D VGG11 outputs 5 numbers to select 5 axial slices. Subsequently, 2D VGG11 predicts 3 scores for each slice.

The 3D CT dataset consisted of 227 SSc patients, with a wide range of disease severity. All patients were scanned with the same CT scanner without contrast enhancement. All 3D CT scans were resized to a fixed size of 256×256×256 with 1.2×1.2×1.2 mm voxel size. The craniocaudal world positions of five levels and the Goh scores for each level were manually annotated by a radiologist (LK, over 20-year experience in chest CT) and a rheumatologist (AS, over 5-year experience) in consensus. In addition, a subset of randomly selected 16 patients was scored by LK and AS twice independently to estimate intra- and inter-observer variation. Four-fold cross validation was applied in both networks.

### 2.1 Level selection

We converted the world positions of the five levels to relative slice numbers in the resampled 3D CT scans (the bottom slice was regarded number 0). The relative slice numbers were used as the reference during training of 3D VGG11. To increase the diversity of training samples, random crops with a fixed size of 256×256×192 (ordered by *xyz*) were applied on-the-fly. The patch always covered all five levels and was also suitable for the GPU memory constrains.

### 2.2 Goh score prediction

The 2D slices at five levels from 227 patients were used for Goh score prediction. The manually annotated scores were rounded to the nearest five percent (shown in Figure 2), following the protocol by Goh[1].

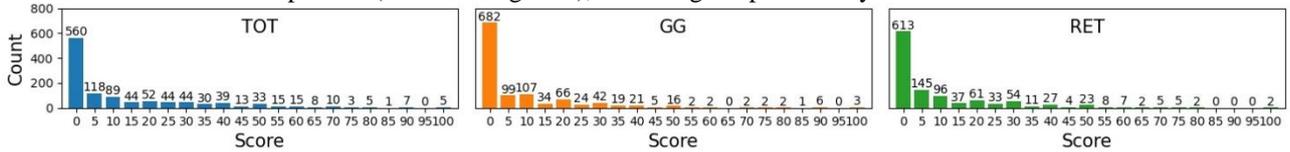

Figure 2. Label distribution of 1135 (5 × 227) slices for three patterns.

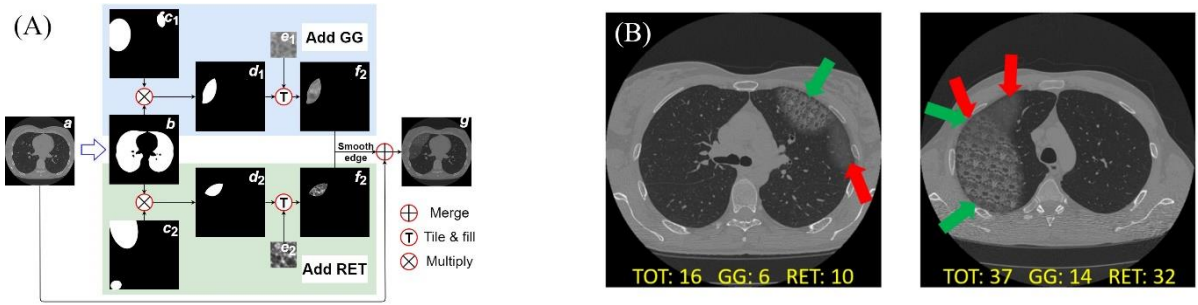

Figure 3. A) Flow chart to synthetize images; B) two synthetic examples (left: without overlap, right: with overlap), where the arrows indicate GG (red) and RET (green) areas. Goh scores of three patterns are at the bottom of each image.

To alleviate the adverse effect of the label imbalance in the existing patient database, in which most labelled slices correspond to 0% scores (healthy) and very few to very high scores (Figure 2), we proposed two techniques: *balanced sampling* and *data synthesis*. *Balanced sampling* resamples the training images with probability inversely proportional to the ratio of each score. It can make sure the distribution of valuable scores is balanced, but it also introduces a large number of repeated images. With the *data synthesis* approach, new images are simulated with new scores (see Figure 3), as follows. First, from selected slices without lesions (*a*), a binary lung mask (*b*) was made[7]. Then two binary masks $c_1$ and $c_2$ were generated by defining up to 3 ellipses with random centers, orientations and axes lengths, which were multiplied by the lung mask to get $d_1$ and $d_2$. The two masks were then filled with GG ($e_1$) and RET ($e_2$), respectively, to get images $f_1$ and $f_2$. Two small patches $e_1$ and $e_2$, which fully contained two different patterns, were manually cropped from diseased slices in advance. Finally, $f_1$ and $f_2$ were inserted into the healthy slice $a$ to obtain the synthetic image $h$. In this process, edges were smoothed to avoid introducing high intensity gradients. The synthetic Goh scores were calculated by the ratio of the areas of the different patterns to the whole lung area. The whole synthesis was applied on-the-fly with a probability of 0.5.

## 3. EXPERIMENTAL RESULTS AND DISCUSSION

The 3D VGG11 was trained on 3D CT scans for level selection, and the 2D VGG11 was trained on 2D slices of five levels for Goh score prediction. Both networks were executed on a machine equipped with a GPU NVIDIA GeForce RTX 2080TI with 11 GB memory. The source code is available at https://github.com/Jingnan-Jia/ssc_scoring.

To have a complete evaluation of networks, apart from mean absolute error (MAE), Weighted Kappa (WK) was calculated using *scikit-learn*; ICC estimates were calculated using the *pingouin*[8] package in Python based on a single-rating, absolute-agreement, 2-way random-effects model; Wilcoxon signed rank test was used to test the significance of the differences between two pairs of results. All metrics were calculated based on 4-fold cross validation.

### 3.1 Level selection

Table 1. Slice prediction results of each level. MAE values are followed by standard deviations (STD).

| Level | 1.Origin great vessels | 2.Main Carina | 3.Venous Confluence | 4.Halfway | 5.Right hemi-diaphragm |
|---|---|---|---|---|---|
| **MAE (#slices)** | 4.21 (3.58) | 3.70 (3.17) | 4.23 (3.55) | 2.49 (2.15) | 2.92 (2.46) |
| **ICC** | 0.72 | 0.84 | 0.81 | 0.96 | 0.97 |

The ICC values indicate moderate reliability on the first level, good reliability on the second and third level, and excellent reliability on the last two levels (Table 1). The reason that the first level is more difficult may be because it requires more anatomical knowledge to locate the origin of great vessels. This is illustrated in Figure 4.A, where most of the points outside of overall 95% CI are from the first three levels. Since the mean error of the prediction was only 0.21 slices, there was no systematic bias. Figure 4.B shows that the regression line (slope = 0.97, intercept = 4.09) nearly coincided with the identity line, which shows outstanding agreement between the predictions and labels.

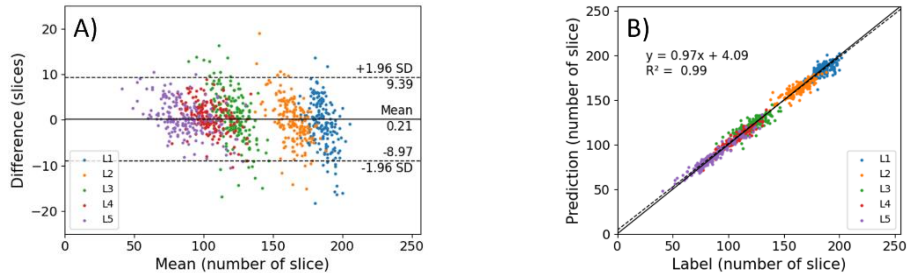

Figure 4. A) Bland-Altman plot and B) Correlation plot of level selection network. In the right plot, dash line is the regression line and solid line is the identify line.

### 3.2 Score prediction

Table 2 compares the validation results of our Goh score prediction networks using different techniques. It shows that both balanced sampling and synthetic data augmentation can help improving performance significantly.

Table 2. Goh score prediction results from 2D Vgg11. ↓: lower is better, ↑: higher is better. †: significantly better than the upper row (P<0.05). All experiments are trained based on pre-trained weights from ImageNet.

| Balanced sampling | Data synthesis | TOT | | | GG | | | RET | | |
|---|---|---|---|---|---|---|---|---|---|---|
| | | MAE↓ | WK↑ | ICC↑ | MAE↓ | WK↑ | ICC↑ | MAE↓ | WK↑ | ICC↑ |
| - | - | 7.85 (10.71) | 0.53 | 0.72 | 5.96 (10.03) | 0.45 | 0.60 | 5.81 (8.42) | 0.53 | 0.72 |
| √ | - | 6.87 (9.61)† | 0.59 | 0.77 | 5.09 (9.50)† | 0.54 | 0.66 | 5.11 (7.45)† | 0.59 | 0.78 |
| √ | √ | **5.90 (8.77)†** | **0.66** | **0.83** | **4.66 (8.83)†** | **0.58** | **0.71** | **4.49 (6.70)†** | **0.65** | **0.84** |

### 3.3 Comparison with human experts

For predicting consensus TOT, our method was close to the first rating by observer AS (AS_T1), but LK was closer to the consensus than our method (Table 3). For GG the model had a fair agreement with the consensus, while the observers have moderate agreement, and for RET the model's agreement was moderate, but moderate/substantial for observers.

Table 3. Comparison between human performance and our method (with pre-trained weights, data synthesis and balanced sampling) in a subset of 16 patients. T1 and T2 denote the first and second observations.

|  | TOT | | | GG | | | RET | | |
| --- | --- | --- | --- | --- | --- | --- | --- | --- | --- |
|  | MAE↓ | WK↑ | ICC↑ | MAE↓ | WK↑ | ICC↑ | MAE↓ | WK↑ | ICC↑ |
| AS_T1 | 7.06 (7.97) | 0.51 | 0.73 | 5.63 (6.14) | 0.44 | 0.68 | 4.94 (7.60) | 0.56 | 0.76 |
| AS_T2 | 6.19 (6.63) | 0.58 | 0.82 | 5.38 (7.53) | 0.46 | 0.59 | 4.75 (7.24) | 0.58 | 0.78 |
| LK_T1 | 6.56 (7.61) | 0.58 | 0.80 | 5.38 (7.53) | 0.48 | 0.63 | 4.63 (6.16) | 0.61 | **0.84** |
| LK_T2 | **4.94 (6.45)** | **0.67** | **0.86** | **4.94 (5.99)** | **0.55** | **0.75** | **4.19 (6.96)** | **0.63** | 0.80 |
| Our method | 8.13 (8.15) | 0.49 | 0.73 | 6.94 (8.46) | 0.33 | 0.47 | 4.81 (6.35) | 0.60 | 0.83 |

## 4. CONCLUSION

We proposed a framework to automatically predict Goh score for lung disease in systemic sclerosis. The synthetic data was introduced to improve the data diversity and alleviate the influence of label unbalance. The framework performed slightly worse than human experts on TOT and GG prediction, but its performance was comparable for RET prediction and has potential to be an objective alternative for visual subjective scoring of systemic sclerosis.

## ACKNOWLEDGEMENT

Computing resources from the Academic Leiden Interdisciplinary Cluster Environment (ALICE) was provided by Leiden University. This work is supported by the China Scholarship Council No.202007720110.